\documentclass
[preprint,aps,superscriptaddress,floats,showkeys,showpacs]{revtex4}%
\usepackage{graphicx}
\usepackage{amsfonts}
\usepackage{amsmath}
\usepackage{amssymb}
\usepackage{graphicx}%
\begin{document}
\title[Optimized 2D structures]{Aperiodic nano-photonic design}
\author{Ioan L. Gheorma}
\affiliation{Department of Electrical Engineering, University of Southern California, Los
Angeles, CA, 90089-2533}
\author{Stephan Haas}
\affiliation{Department of Physics and Astronomy, University of Southern California, Los
Angeles, CA, 90089-0484}
\author{A. F. J. Levi$^{1,}$}
\affiliation{Department of Physics and Astronomy, University of Southern California, Los
Angeles, CA, 90089-0484}
\author{}
\keywords{adaptive algorithm, optimization, aperiodic, photonic crystal, scattering}
\pacs{42.70.Qs, 02.60.Pn, 11.80.La}

\begin{abstract}
The photon scattering properties of aperiodic nano-scale
dielectric structures can be tailored to closely match a desired
response by using adaptive algorithms for device design. We show
that broken symmetry of aperiodic designs provides access to
device functions not available to conventional periodic photonic
crystal structures.

\end{abstract}
\volumeyear{2003}
\volumenumber{xx}
\issuenumber{xx}
\eid{identifier}
\startpage{1}
\endpage{3}
\maketitle


\subsection{Introduction}

The spatial arrangement of nano-scale dielectric scattering
centers embedded in an otherwise uniform medium can strongly
influence propagation of an incident electromagnetic wave.
Exploiting this fact, we iteratively solve an inverse problem
\cite{foot1} to find a spatial arrangement of identical
non-overlapping scattering centers that closely matches a desired,
or target, response. Of course, the efficiency of adaptive
algorithms used to find solutions may become an increasingly
critical issue as the number of scattering centers, $N$,
increases. However, even for modest values of $N$, this method
holds the promise of creating nano-photonic device designs that
outperform conventional approaches based on spatially periodic
photonic crystal (PC) structures.

The configurations considered by us consist of either lossless dielectric rods
in air or circular holes in a dielectric similar to the majority of quasi two
dimensional (2D) PCs reported in the literature by, for example, \cite{bend,
haus}. To confirm the validity of our 2D simulations we compare the optimized
configuration with full three dimensional (3D) simulations.

\subsection{Calculation of the scattered field}

The optimization process is an iterative procedure, in which the scattered
field from a trial configuration of cylindrical rods or circular holes is
compared with that of the target. We note that computation of the scattered
field is sometimes referred to as the forward problem.

The electromagnetic field solver used is based on the analytical solution of
the Helmholtz equation by separation of variables in polar coordinates. A
typical problem we consider is a set of $N$ long, parallel, lossless circular
dielectric rods distributed in a uniform medium and illuminated by an
electromagnetic wave perpendicular to the axis of the cylinders. The natural
geometry of the system led us to use a 2D electromagnetic field solver. When
analyzing scattering from one cylinder, the solution, expressed as a
Fourier-Bessel series, is found immediately by imposing the continuity of the
electric and magnetic field components at the rod surface. However, when
studying scattering from two or more cylinders, multiple scattering gives an
additional linear system that has to be solved in order to find the
Fourier-Bessel coefficients \cite{maystre1}. For a given number of Bessel
functions used this linear system has a reduced form which is conveniently
described using the scattering matrix method
\cite{maystre2,maystre3,maystre4,yonekura,Ye}. The input wave can, in
principle, be of arbitrary shape so long as it is expressed as a
Fourier-Bessel series. In our simulations the input beam is a gaussian and
both TE and TM polarizations are considered. The calculated Fourier-Bessel
coefficients for the gaussian beam conform \cite{kozaki, pier}. Additional
details on the electromagnetic solver used may be found in Appendix \ref{em}.

\subsection{Optimization}

For the results presented in this paper, the optimization method used is the
guided random walk. The positions of individual cylinders are randomly changed
by a small amount, the modified distribution is simulated, and if the result
is closer to the target function, such that the error defined in Appendix B is
decreased, then the new configuration is accepted, otherwise it is rejected.
The actual implementation of the adaptive algorithm also includes other types
of collective motion such as moving more than one cylinder per iteration, and
moving or rotating all the cylinders.

In general the target function may be any electromagnetic distribution. A
typical target function might involve redirecting and reshaping the input
beam. More complex situations involving optically active materials are not
considered in this paper. Because the target function can specify the field
distribution of a guided mode, it is possible to create mode converters to
interface between a fiber or ridge waveguide and a PC waveguide. In this paper
we report on redirecting and reshaping an electromagnetic beam using optimized
aperiodic nano-structures.

\subsection{Results}

To illustrate our approach, we considered an input beam of
gaussian profile scattered by an angle of $45^o$. The target
functions we consider are a top hat distribution of the optical
intensity with respect to the scattering angle and a cosine
squared ($cos^{2}$) distribution of the intensity. We note that a
modal field distribution target function requires amplitude and
phase to be specified.

As an initial demonstration we chose the top-hat intensity function because it
is difficult to achieve (even approximately) in conventional optical systems
and serves to show the capabilities of our adaptive algorithm. Application of
such an intensity distribution includes guaranteeing uniform illumination of
the active area of a photodetector. On the other hand the cos$^{2}$ intensity
distribution target approximates the transverse spatial mode intensity
typically found in a waveguide and so could be considered a first step toward
design of a waveguide coupler.

\subsubsection{Top-hat target function}

Optimization of a top-hat intensity distribution target function
is performed starting from a configuration of $N=56$ dielectric
rods (represented in Fig. \ref{initial1}(a) by the small circles),
each having refractive index $n=1.5$, and diameter $d=0.4$ $\mu$m.
The medium surrounding the rods is air and the structure is
illuminated by a TM polarized (electric field along the \textit{z}
direction) gaussian beam of width $2\sigma=4$ $\mu$m, wavelength
$\lambda=1$ $\mu$m, and propagating along the positive $x$
direction (from left to right in Fig. \ref{initial1}(a)). The
initial configuration of the rods and intensity distribution are
illustrated in Fig. \ref{initial1}(a), where the arrows represent
(in arbitrary units) the real part of the Poyting vectors. The
target function window (represented in Fig. \ref{initial1}(a) by a
missing arc in the $7$ $\mu$m radius observation circle) extends
from $30^{o}$ to $60^{o}$. Fig. \ref{initial1} (b) shows
normalized intensity (real part of the normal component of total
field Poynting vectors directed outwards) as a function of angle
on a radius of $7$ $\mu$m from the center of the symmetric array.\
Clearly, for the initial configuration, the overlap with the
top-hat target function (broken line) is poor.

Fig. \ref{final1}(a) shows the spatial distribution of the $N=56$
rods with the real part of Poynting vectors after 9700 iterations
of the adaptive search algorithm. Fig. \ref{final1}(b) shows the
corresponding angular distribution of the intensity. In Fig.
\ref{final1}(c) the distribution of the electric field (relative
magnitude, with 1 corresponding to the maximum magnitude in the
incident gaussian beam) is shown and in Fig. \ref{final1}(d) the
relative error versus number of iterations (the errors are
normalized with respect to the initial value) are represented.
Note that for this system with $2\times N$\ positional degrees of
freedom, the error is not saturated even after 9,700 iterations
(Fig. \ref{final1}(d)) and the structure can be further optimized
by performing additional iterations. The error is computed using
the method described in Appendix \ref{err} with exponent
$\gamma=2$ on the observation circle of radius $7$ $\mu$m.

For comparison, in Fig. \ref{PC_comp} we show the results of using a PC with
the same number of dielectric rods, $N=56$. Clearly, the number of rods is not
sufficient to redirect all the beam in the 45$^{o}$ ($\pm$15$^{o}$) direction
and the error with respect to the target function is unacceptably large. The
spatial symmetry of the PC excludes access to the top-hat target function. It
is only by breaking the symmetry that one may acquire the target. In general,
broken symmetry enables functionality.

\subsubsection{Cosine squared target function}

For a $cos^{2}$ target function we chose a different situation in
which there are $N=26$ lower index cylinders (SiO$_{2},$ $n=1.45$)
embedded in a higher index material (Si, $n=3.5$). The 3D
equivalent of the modelled situation is a Si slab with cylindrical
perforations embedded in SiO$_{2}$. The incident wave is a TE
polarized (magnetic field along the \textit{z} direction) gaussian beam of
width $2\sigma=1.5$ $\mu$m and wavelength $\lambda=1.5$ $\mu$m. In
a similar manner to the previous example, the initial
configuration and intensity distribution are illustrated in Fig.
\ref{initial2}. The back scattered intensity in the input region
is calculated as the difference between the incident field
Poynting vector and total field Poynting vector. In Fig.
\ref{final2} we show the optimized configuration, Poynting
vectors, relative error versus number of iterations, and relative
magnitude of the magnetic field (with 1 corresponding to the
maximum magnitude in the original gaussian beam). This time the
position of the $N=26$ cylinders could be changed and so could
their diameters $d$, giving a total number of degrees of freedom
that is three times the number of cylinders $N$. The values of the
diameters are constrained to the range $0.2\leq d\leq0.5$
$\mu$m$.$ The error is computed using the metric from Appendix
\ref{err} with exponent $\gamma=1$ on a $6$ $\mu$m radius circle.
Notice that due to the small number of cylinders the actual number
of degrees of freedom is smaller (78) relative to the previous
calculations with a top-hat target function (112) and the
optimization saturates after a comparatively small number of
iterations.

\subsubsection{Computing resources}

Our previous optimization work on 1D problems \cite{yu} required relatively
insignificant computational resources. Optimization algorithms using 2D
electromagnetic solvers are always more compute intensive than those for 1D
structures. Interesting simulations in 3D are even more compute intensive and
often require parallel computing. Each of the simulations (top-hat and
$cos^{2}$ target) discussed in this paper took 4 days to complete (9,700
iterations in the first case and 7,600 of the second) using a Pentium IV
processor with a 3 GHz clock frequency, 533 MHz memory bus, and 1 GB RDRAM.

The compute time for the forward problem solver is dominated by the solution
of a linear system with a full matrix of complex numbers and thus is strongly
dependent on the number of cylinders $N$ considered in the problem and the
number of Bessel functions $N_{b}$ used. The size of the system is
$N_{m}=N(2N_{b}+1)$ and the solver routine (iteratively corrected LU
decomposition \cite{NR}) time is proportional to the cube of the matrix size
($O(N_{m}^{3})$). When using an incident plane wave or a cylindrical wave the
number of Bessel functions needed is very small, however, an appropriately
accurate approximation of the gaussian beam shape requires a large number of
Bessel functions and a corresponding increase in compute time.

\subsubsection{Comparison with 3D Simulation}

For practical purposes, a comparison with 3D electromagnetic simulations
serves to confirm the accuracy of solutions obtained with the 2D simulator.
Also, since 3D simulations are much more time consuming than 2D ones, one
might adopt the 2D optimization result as a starting point for a limited
number of 3D iterations which are then used to further refine optimization.

In this paper we compare results of our 2D electromagnetic simulations with 3D
Finite Integration Technique (FIT) simulations obtained using a commercially
available package, CST Microwave Studio \cite{CST}.

In a realistic structure one might anticipate the infinitely long
cylindrical hole structure used in the $cos^{2}$ target simulation
to be replaced with SiO$_{2}$ filled holes in a Si slab itself
embedded in SiO$_{2}$. The input beam might be launched into this
slab from a ridge waveguide. In our simulations, the wavelength of
the light is $\lambda=1.5$ $\mu$m and the polarization is TE. The
Si slab is $0.6$ $\mu$m thick and the effective index of the
fundamental mode of this slab waveguide is the same as the index
of the material surrounding the cylinders in the 2D simulation.
The diameter of the holes is $d=0.4$ $\mu$m and the mode size of
the ridge waveguide (Fig. \ref{ribmode}) is approximately 1.5
$\mu$m and so the same as the width of the gaussian beam used for
the 2D simulations. The differences between the two simulations
(Fig.
 \ref{ribmode}(b) for the 3D and Fig. \ref{ribmode}(d) for the 2D simulation results) are
primarily due to the nonuniformity of the field and structure in
the \textit{z} direction but also because of the difference
between the gaussian beam and the ridge mode and the fundamently 3D
discontinuity between the ridge waveguide and the slab. These
differences translate into a 30\% decrease in peak intensity and
25\% decrease in the total power directed in the desired direction
for the 3D simulation compared to the 2D one (Fig. \ref{ribmode}
(c)).

\subsubsection{Sensitivity analysis}

We analyzed the influence of small changes in the wavelength of
the incident beam on error. We took the optimized configuration
from the $cos^{2}$ target example (Fig. \ref{final2}(a)) and
simulated the effect of changing the frequency of the input light.
The relative deviation from the minimum error is shown in Fig.
\ref{modulated} as a function of the electromagnetic wave
frequency shift. Frequency variations of $\Delta f=200$ GHz only
change the error by 0.3\% of its minimum value. Thus an optical
beam centered at wavelength $\lambda=1.5$ $\mu$m modulated at very high speed
will behave essentially as the simulated monochromatic wave at
$\lambda=1.5$ $\mu$m (200 THz). Even a 1 THz deviation in
frequency changes the error by only 6 -- 7 \%.

Another important issue is the sensitivity of the aperiodic nano-photonic
design to variations and errors introduced by the fabrication process. As an
initial study we explored the sensitivity of the design to changes in the
position of the cylinders. Other parameters that are influenced by fabrication
processes such as the diameter and detailed shape of the cylinders, index of
refraction, direction of the input ridge waveguide with respect to the
cylinders, etc., are not analyzed in this paper.

To estimate the sensitivity to position, for the same $cos^2$
target function example, we use the following method. First, one
by one the cylinders are displaced by a small fixed distance
$\Delta$ in the positive and negative $x$ and $y$ directions and
the change in error is evaluated for each cylinder (the maximum
error created out of the 4 displacements). The displacement values
used are small compared to the initial diameters of the cylinders
($d=400$ nm). Five values for the displacement $\Delta$ are
considered: 10, 20, 30, 40, and 50 nm. Next the 10 displaced
cylinders with the greatest influence on the error are selected.
These cylinders, as expected, are located in regions associated
with high field intensity.

The selected cylinders are all individually moved randomly by the
same step size in the positive and negative $x$ and $y$ directions
(each movement has the same 1/4 probability). The movement of each
cylinder is independent of the movements of the other selected
cylinders. This way we generate a number of slightly modified
configurations. The errors for 10 of these modified configurations
are computed for each displacement value and a combined plot
generated (Fig. \ref{histo}).

We chose this very simplified method for estimating the sensitivity to
position because a more complete approach would involve independent
displacements of each cylinder in completely random directions and with
variable distances and hence give rise to significant computational effort.
The plot in Fig. \ref{histo}, suggests that 10 nm precision in fabrication
might be needed to ensure less than 10\% increase in minimum error for devices
operating at wavelength $\lambda=1.5$ $\mu$m.

\subsubsection{Comparison with photonic crystal inspired devices}

In recent years, spatially periodic dielectric structures have
been studied and applied to both optics and microwaves
\cite{PCbook, mwPC, mwPC2}. It has been shown that introduction of
point and line defects in PCs can be used to filter, demultiplex,
and guide electromagnetic waves \cite{haus, bend,
demux,chaneldrop}. However, there are numerous unresolved design
issues with PC inspired devices that limit prospects of adoption
as practical components. For example, when coupling between
standard fibers or waveguides and PC waveguides, the back
reflection is either unacceptably large \cite{coupling2pc3} or
requires use of relatively large coupling regions
\cite{coupling2pc,coupling2pc2}. As another example, it is well
known that the necessarily finite size of the PC can have a
dramatic and detrimental impact on device performance
\cite{finitepc}. Solutions to these and similar problems are
stymied by the limited number of degrees of freedom inherent to PC
design. Rather than struggle for solutions within the constraints
of spatial symmetry imposed by PC structures, our approach is
based on breaking the underlying spatial symmetries and thereby
exposing larger numbers of degrees of freedom with which to design
and optimize nano-photonic devices.

Our experience so far indicates that optimization of such systems
is best achieved using numerical adaptive design techniques. This
is because solutions, such as that illustrated in Fig.
\ref{final2}, have such a high degree of broken symmetry it is
unlikely analytic methods or conventional intuition could usefully
be applied.

\subsubsection{Conclusions}

In this paper we have shown that aperiodic nano-photonic dielectric structures
designed using adaptive algorithms can be tailored to closely match desired
electromagnetic transmission and scattering properties. It is the broken
symmetry of the structure that allows more degrees of freedom and the
possibility of better optimization compared to symmetric photonic crystal structures.

In general the frequency response and spatial response of this system can have
very complicated forms. It is the large number of degrees of freedom that
allow us to tailor the response to the desired target. The configuration space
is even more complex if materials with optical loss or gain are considered.

\begin{acknowledgments}
We thank P. B. Littlewood for helpful comments. This work is supported in part
by DARPA.
\end{acknowledgments}

\appendix

\section{Electromagnetic solver}

\label{em}

For brevity we will only discuss TM polarized electromagnetic waves. For the
TE case the equations are similar with $H$ replacing $E$. The total field is
written as the sum of the incident field $E_{inc}$ and the field $E_{sc}$
scattered from the cylinders \ $E=E_{inc}+\sum_{i=1}^{N}E_{sc}^{i}$. The
actual incident field on a cylinder labled with index $j$ is $E_{inc}%
^{j}=E_{inc}+\sum_{i\neq j}^{N}E_{sc}^{i}$ \cite{maystre1,Ye}. We are
interested in solving the Helmholtz equation for the total field,\ $\nabla
^{2}E+k^{2}E=0$, where $k=k_{0}$ in the region outside the cylinders and
$k=k_{1}$ inside the cylinders. The method used to solve this equation is
separation of variables in polar coordinates. Of course, the solution of the
homogeneous Helmholtz equation in polar coordinates is well known. All field
quantities may be written in the form of Fourier-Bessel series with the
coefficients $\alpha_{m}$ and $\beta_{m}$ determined from the boundary
conditions. \ Hence, $E=\sum_{m=-\infty}^{\infty}\alpha_{m}Z_{m}%
(k\rho)e^{im\theta}+\sum_{m=-\infty}^{\infty}\beta_{m}\widetilde{Z}_{m}%
(k\rho)e^{im\theta}$, where $Z_{m}$ and $\widetilde{Z}_{m}$\ are two conjugate
cylindrical functions. \ These functions are either the first order Bessel
functions $J_{m}$ and $Y_{m}$ or the second order Bessel functions (Hankel
functions) $H_{m}^{(1)}$ and $H_{m}^{(2)}$. The pair chosen depends on the
boundary conditions. Outside the cylinders the asymptotic behavior determines
which functions are used. Since only the $H_{m}^{(2)}$ function has the
behavior of an out-propagating cylindrical wave, the field of the scattered
wave $E_{sc}$ has to be written using only $H_{m}^{(2)}$ of the \{$H_{m}%
^{(1)}$, $H_{m}^{(2)}$\} pair. Hence the scattered field is $E_{sc}%
=\sum_{m=-\infty}^{\infty}b_{m}H_{m}^{(2)}(k_{0}\rho)e^{im\theta}$. Inside the
cylinders we have to choose the \{$J_{m}$, $Y_{m}$\} pair because the Hankel
functions are both singular at the origin. This behavior comes from the
$Y_{m}$ part of the Hankel function ($H_{m}^{(1)}=J_{m}+iY_{m}$, $H_{m}%
^{(2)}=J_{m}-iY_{m})$ and thus only the $J_{m}$ functions can be kept for the
interior. In this case the total internal field is $E_{tot}^{int}%
=\sum_{m=-\infty}^{\infty}a_{m}J_{m}(k_{1}\rho)e^{im\theta}$.

After expressing the incident field in polar coordinates, i.e. in a
Fourier-Bessel series, we can write the electric and magnetic field continuity
conditions at the boundaries of each cylinder. This gives us a system of
equations with the unknowns being the Fourier-Bessel coefficients of the
scattered field outside the cylinders and total field inside the cylinders.
The system can be simplified by using the relationship between the
Bessel-Fourier coefficients of a field incident on a cylinder and the
coefficients for the scattered and internal fields
\cite{maystre2,maystre3,maystre4,yonekura}.

\section{Error function}

\label{err}

The optimization is based on the minimization of a functional,
which is defined as the residual error between the calculated
angular distribution of the normal component of the Poynting
vector $\mathbf{S}$ and a distribution expressed as a target
function. This error function is computed starting from the
difference in intensity between the target and the result. In 2D
the error is calculated along an observation line (often a circle
around the group of cylinders). This line is divided into small
portions and the normal component of the S vector is calculated in
the center of each segment.

Let the target function $T(\alpha)$ be the angular distribution of intensity
exiting the circular observation region and $S_{n}(\alpha)$ the normal
component of the real part of the Poynting vector ($S_{n}(\alpha
)=\mathbf{S}(\alpha)\cdot\mathbf{n(}\alpha\mathbf{)}$ where $\mathbf{n}$ is
the normal unit vector). In this space of functions defined on $[0^{o}$,
$360^{o}]$ and having real values we can define a "distance" $D$ between
result $S_{n}(\alpha)$\ and target $T(\alpha)$\ as $D=\frac{1}{2\pi}\int
_{0}^{2\pi}\left\vert S_{n}(\alpha)-T(\alpha)\right\vert ^{\gamma}d\alpha$. To
properly evaluate the difference between the target and result the functions
$T$ and $S$ and have to be similarly normalized.

In general, the exponent $\gamma$ can take any value. Choosing $\gamma
=1$\ assures that each improvement is considered with the same weight. When
choosing $\gamma>1$\ improvements made in regions where the target and the
results are very different influence the integral $D$ more than a few smaller
improvements in other regions. This means that $\gamma=1$ tends to ensure an
uniform convergence while $\gamma>1$ favors reduction of major differences
between target and result. The greater the numerical value of $\gamma$, the
more important this effect becomes, while for $0<\gamma<1$\ the effect is
reversed. And finally, a negative exponent $\gamma$ tends to push the solution
further away from the target, in a manner which depends on the numerical value
of $\gamma$.

One could use different exponents for different stages of the iterative
process. For example $\gamma=1$ could be used at the beginning of the
convergence procedure to avoid local minima. Later, the value of $\gamma$
could be increased to accelerate convergence towards a minimum.\ Furthermore,
if one decides that this minimum is not sufficiently close to the target
function, application of a negative exponent would divert the iterations from
this local minimum to some intermediate point were $\gamma>1$ iterations could
be restarted in the search for a better minimum.

We use the real part of the Poynting vector in our calculation of
$S_{n}(\alpha)$. For most scattering directions this is the Poynting vector of
the total field, with the exception of the input beam region where we use the
scattered field in the first example (the top-hat target function) and the
difference between the Poynting vector of the input field and total field in
the second example (the $cos^{2}$ target function).

The distances, or errors, \textit{D} are computed numerically by dividing the
observation circle into very small and equal portions and the integral
replaced by a sum over these portions \ $D^{\ast}=\sum_{i=0}^{Np}\left\vert
\frac{S_{n}(\alpha_{i})}{N_{s}}-\frac{T(\alpha_{i})}{N_{t}}\right\vert
^{\gamma}$ \ where $N_{s}$ and $N_{t}$ are normalization factors. Different
normalization methods can be used such as normalization to the maximum value,
normalization to the sum of all values, or normalization to the sum of the
squares. Since in our case the functions involve intensity we have chosen the
normalization to be the sum \textit{i.e.} the total power.

Although not used in this paper, it is worth mentioning that when optimizing
for a modal shape distribution (both amplitude and phase) a better choice for
error function would be the overlap integral between the actual field
distribution and the desired modal field distribution. In this situation the
target would be maximized.

\begin{figure}[ptbh]
\begin{center}
\end{center}
\caption{Starting configuration for the top-hat target function
example. 56 dielectric rods with diameter $d= 0.4$ $\mu$m, index
$n=1.5$, in air. The incident beam, which propagates along the
\textit{x} axis (left to right), is a TM polarized gaussian beam
with beamwidth $2\sigma=4$ $\mu$m, and wavelength $\lambda=1$
$\mu$m. (a) Positions of the cylinders and distribution of the
Poynting vector field. The observation circle, has a radius of 7
$\mu$m and it is represented without the target window between 30
and 60 degrees. (b) Computed angular intensity distribution
(continuous line) at radius 7 $\mu$m
compared with the target distribution (dashed line).}%
\label{initial1}%
\end{figure}

\begin{figure}[ptbh]
\begin{center}
\end{center}
\caption{Optimized configuration for the top-hat target function
example. 56 dielectric rods with diameter, $d= 0.4$ $\mu$m, index
$n=1.5$, in air. TM polarization gaussian beam with incident
beamwidth $2\sigma=4$ $\mu$m, and wavelength $\lambda=1$ $\mu$m. (a)
Positions of the cylinders and distribution of the Poynting vector
field after 9,700 iterations. The observation circle, has a radius
of 7 $\mu$m and it is represented without the target window
between 30 and 60 degrees. (b) Computed angular intensity
distribution after 9,700 iterations (continuous line) at radius 7
$\mu$m compared with the target distribution (dashed line). (c)
Contour plot of the electric field magnitude in
relative units after 9,700 iterations. (d) Evolution of the error.}%
\label{final1}%
\end{figure}

\begin{figure}[ptbh]
\begin{center}
\end{center}
\caption{Comparison with a periodic structure (PC) with 56
dielectric rods ($n=1.5$) in air. The lattice constant and the
angle are chosen so that for the incident wavelength
$\lambda=1\mu$m the Bragg diffraction condition is satisfied for 45$^{o}.$
(a) Positions of the cylinders and distribution of the Poynting
vector field. The incident beam is propagating along the
\textit{x} axis (left to right). The observation circle, has a
radius of 7 $\mu$m and it is represented without the target window
between 30 and 60 degrees. (b) Computed angular intensity
distribution (continuous line) at radius 7 $\mu$m compared with
the target distribution (dashed line).}%
\label{PC_comp}%
\end{figure}

\begin{figure}[ptbh]
\begin{center}
\end{center}
\caption{Initial configuration for the cosine squared target function example.
Si ($n=3.5$) with 26 cylindrical holes filled with SiO$_{2}$, TM gaussian beam
incidence, beamwidth $2\sigma=1.5$ $\mu$m, and wavelength $\lambda=1.5$
$\mu$m. (a) Positions of the cylinders and distribution of the Poynting vector
field. The incident beam is propagating along the \textit{x} axis (left to
right). The observation circle, has a radius of 6 $\mu$m and it is represented
without the target window between 30 and 60 degrees. (b) Computed angular
intensity distribution (continuous line) at radius 6 $\mu$m compared with the
target distribution (dashed line). }%
\label{initial2}%
\end{figure}

\begin{figure}[ptbh]
\begin{center}
\end{center}
\caption{Optimized configuration for the cosine squared target
function example. Si ($n=3.5$) having 26 cylindrical holes filled
with SiO$_{2}$, TM gaussian beam incidence, beamwidth $2\sigma=1.5$
$\mu$m, and wavelength $\lambda=1.5$ $\mu$m. (a) Positions of the
cylinders and distribution of the Poynting vector field. The
incident beam is propagating along the \textit{x} axis (left to
right) after 7,000 iterations. The observation circle, has a
radius of 6 $\mu$m and it is represented without the target window
between 30 and 60 degrees. (b) Computed angular intensity
distribution after 7,000 iterations (continuous line) at radius 6
$\mu$m compared with the target distribution (dashed line). (c)
Contour plot of the magnetic field magnitude in relative units
after 7,000 iterations. (d) Evolution of the error as a
fucntion of iteration number.}%
\label{final2}%
\end{figure}

\begin{figure}[ptbh]
\begin{center}

\caption{Comparison between 3D and 2D simulations. (a) The 3D
structure simulated and the modal field of the input ridge
waveguide. The thickness of the Si slab in the 3D simulation is
0.6 $\mu$m and the (horizontal) mode size  of the ridge waveguide
is approximately $w=1.5$ $\mu$m, the diameter of the holes is
$d=0.4$
 $\mu$m. Incident light is TE polarized and has wavelength $\lambda= 1.5 \mu$m.
(b) 3D Simulation results showing the Poynting vectors field in the
middle horizontal section of the Si slab. (c) Comparison between
the angular intensity distribution for the 3D and 2D simulations.
(d) 2D simulation results, the Poynting vectors field.}
\label{ribmode}%
\end{center}
\end{figure}

\begin{figure}[ptbh]
\begin{center}
\end{center}
\caption{Sensitivity analysis with respect to the frequency of the
incident light for the optimized structure from Fig. \ref{final2}.
Percentage change in error when the frequency of the light is modified from the original value $f_0=200$ THz . }%
\label{modulated}%
\end{figure}

\begin{figure}[ptbh]
\begin{center}
\end{center}
\caption{Sensitivity analysis with respect to position for the
optimized structure from Fig. \ref{final2}. The 10 most sensitive
cylinders are randomly moved along the x and y axis by a fixed
length step (horizontal axis 10, 20, 30, 40 and 50 nm). The errors
for 10 different randomly generated
moved distributions are plotted for each step size. }%
\label{histo}%
\end{figure}

\clearpage
\resizebox{15cm}{!}{\includegraphics{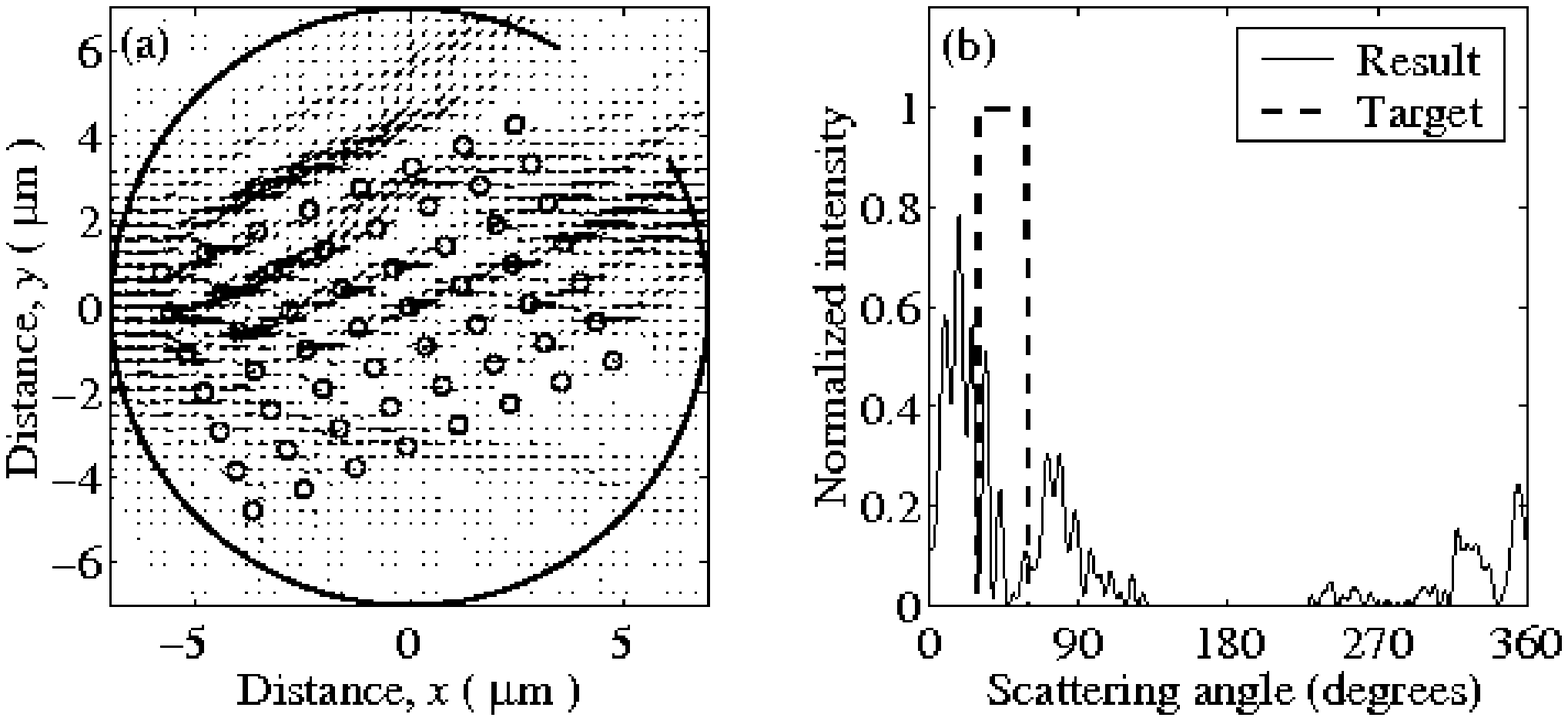}}
\bigskip
\bigskip
\vspace{10 cm}
 \\ Fig. 1
 \\ Gheorma, Haas, Levi

 \clearpage

\resizebox{15cm}{!}{\includegraphics{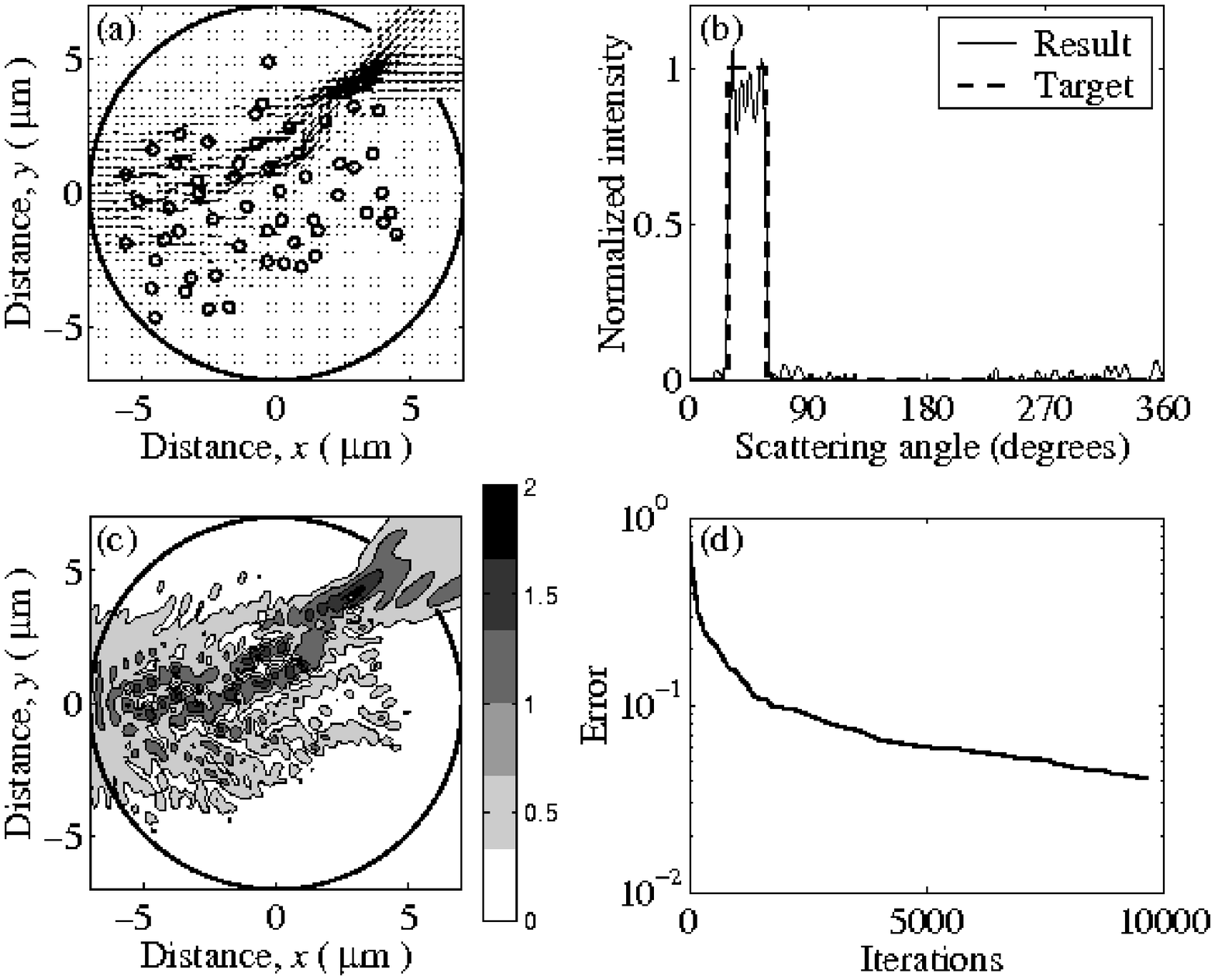}}
\bigskip
\bigskip
\vspace{7 cm}
 \\ Fig. 2
 \\ Gheorma, Haas, Levi

 \clearpage

\resizebox{15cm}{!}{\includegraphics{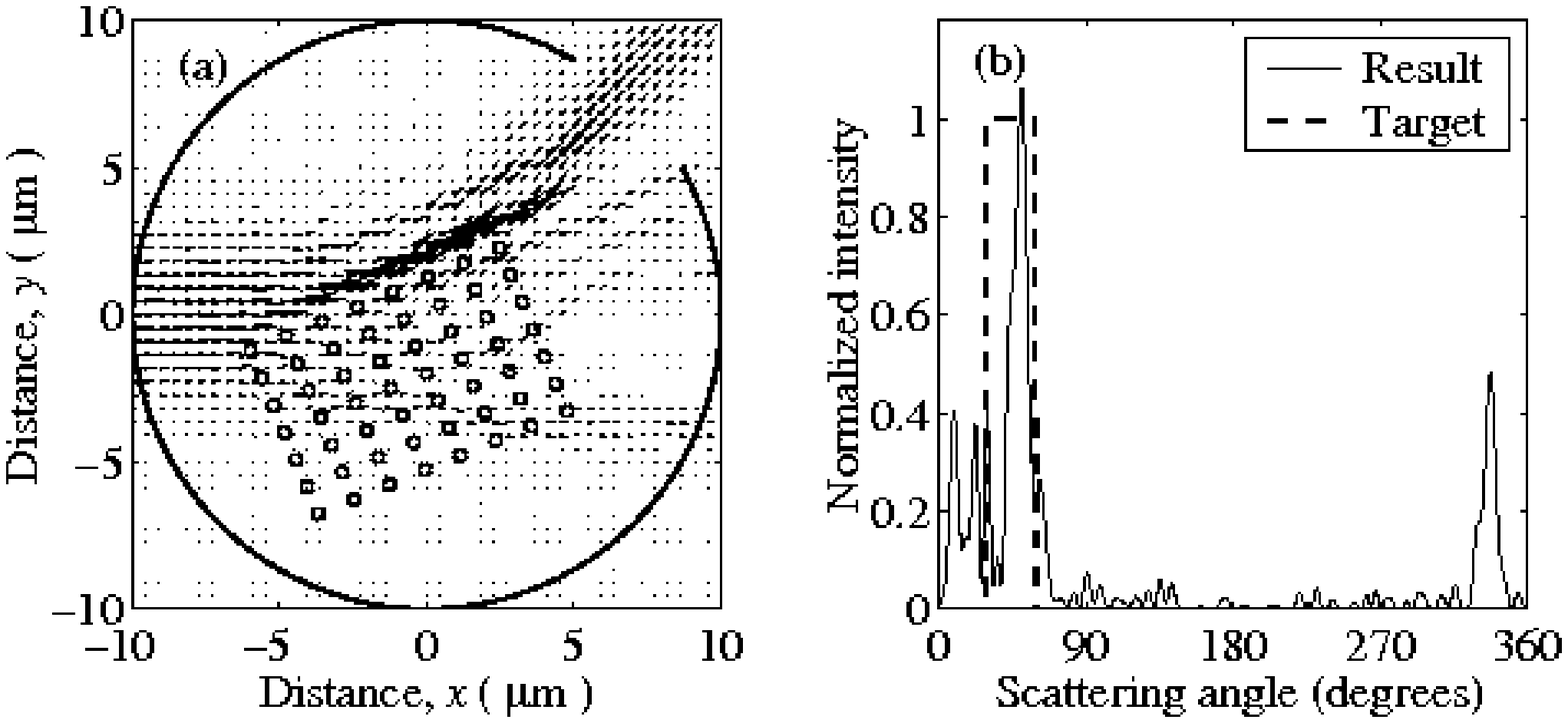}}
\vspace{10 cm}
 \\ Fig. 3
 \\ Gheorma, Haas, Levi

 \clearpage

\resizebox{15cm}{!}{\includegraphics{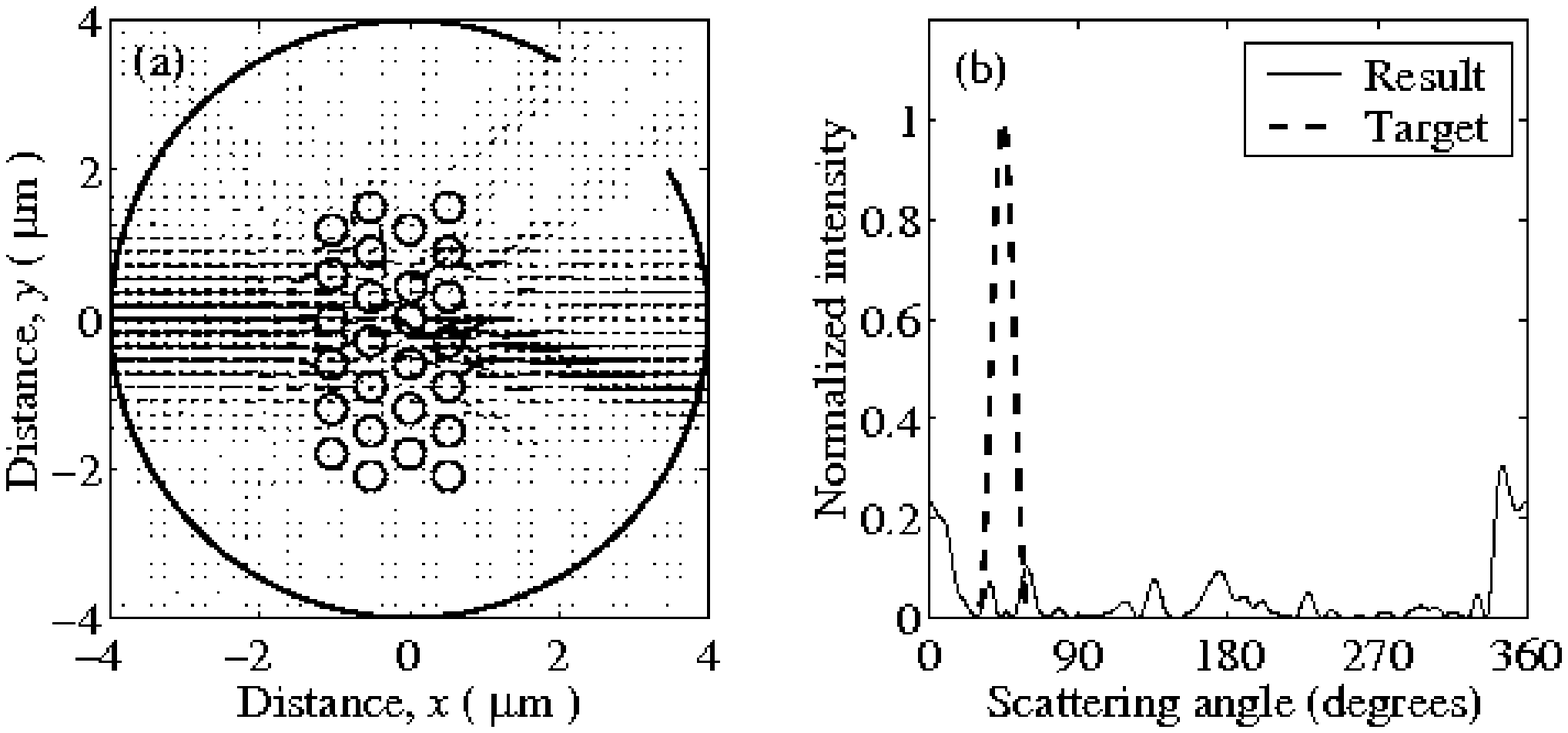}}
\vspace{10 cm}
 \\Fig. 4
 \\ Gheorma, Haas, Levi

 \clearpage

\resizebox{15cm}{!}{\includegraphics{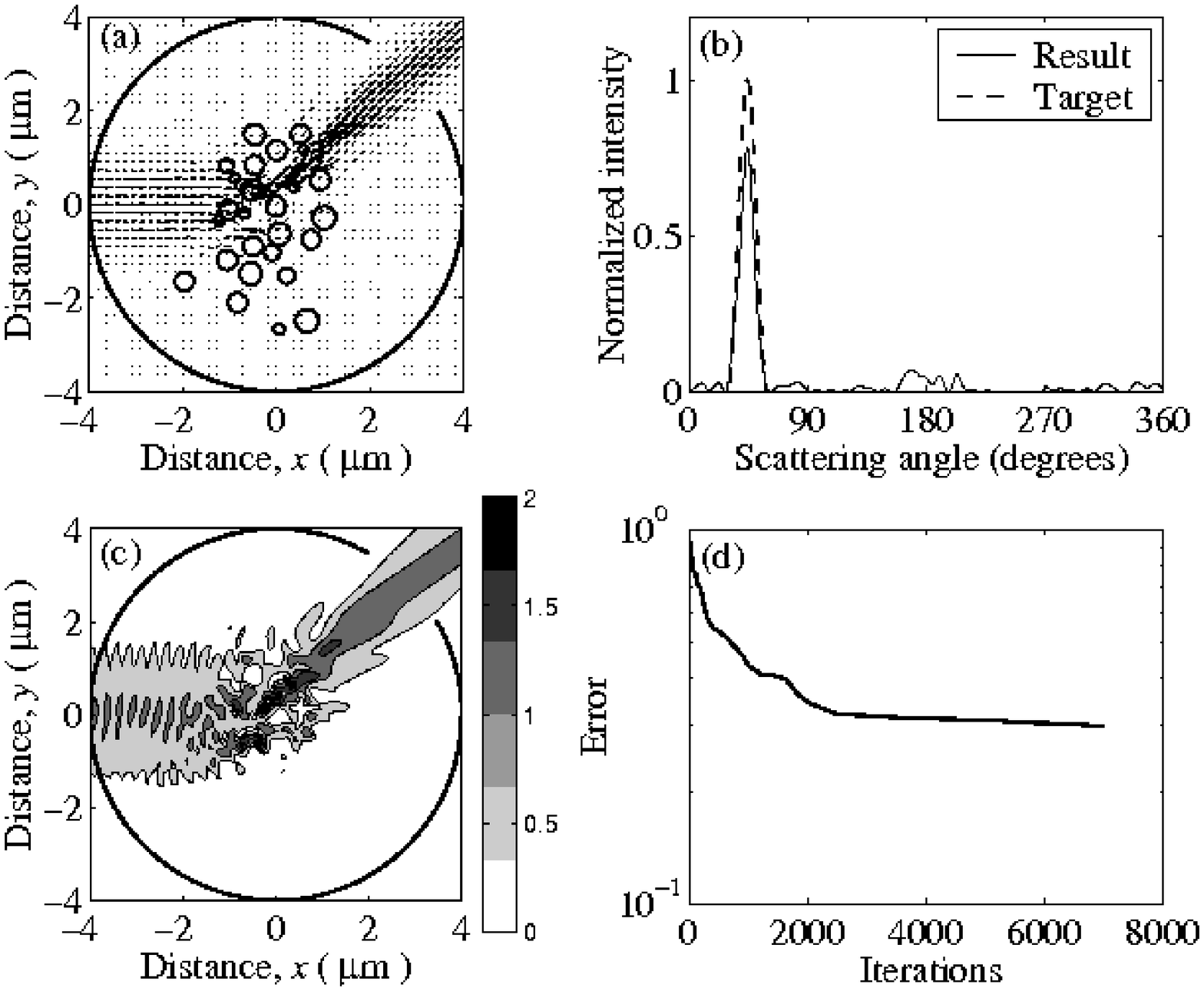}}
\vspace{7 cm}
 \\ Fig. 5
 \\ Gheorma, Haas, Levi

 \clearpage
\resizebox{15cm}{!}{\includegraphics{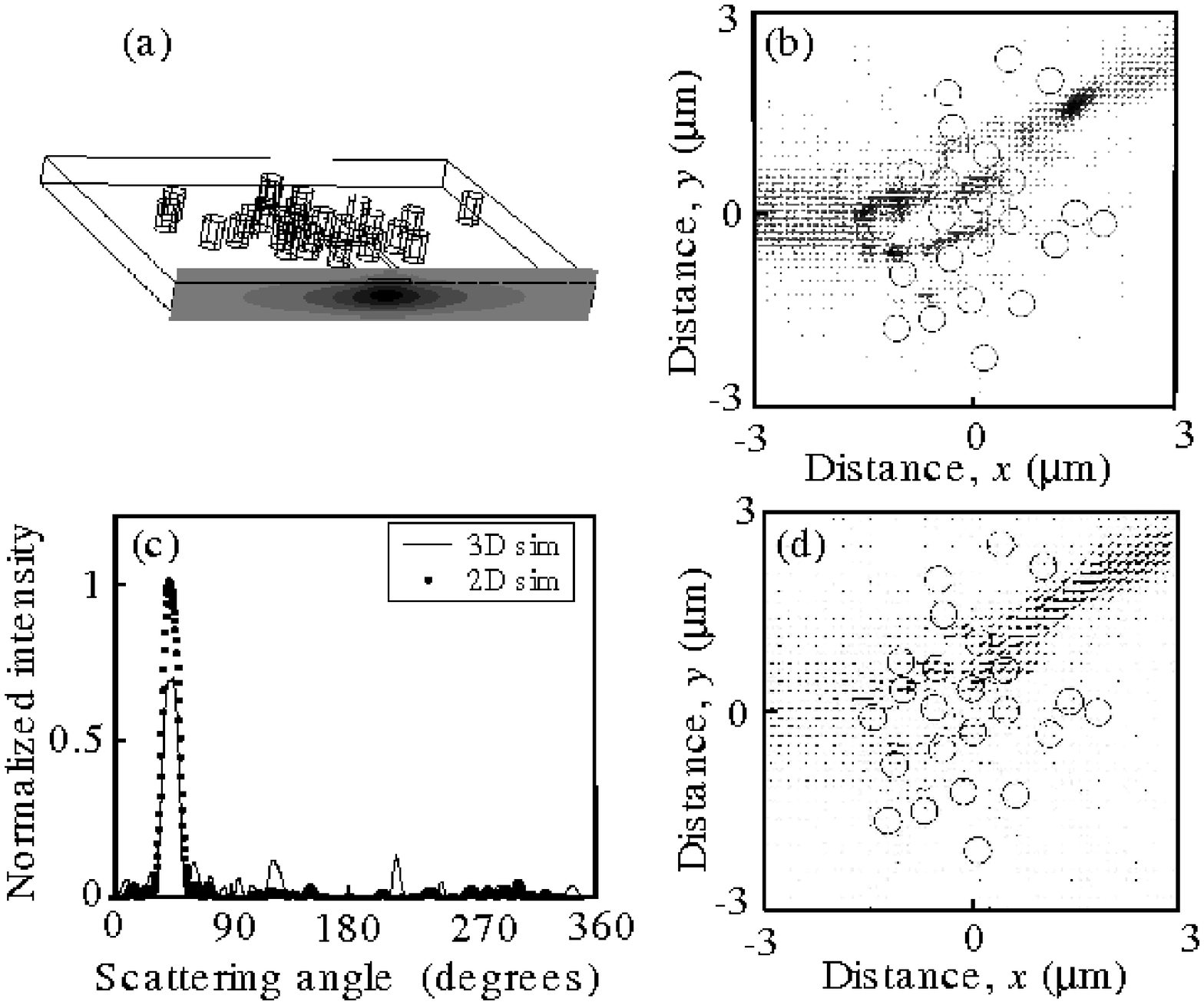}}
\vspace{7 cm}
 \\ Fig. 6
 \\ Gheorma, Haas, Levi

 \clearpage
\resizebox{15cm}{!}{\includegraphics{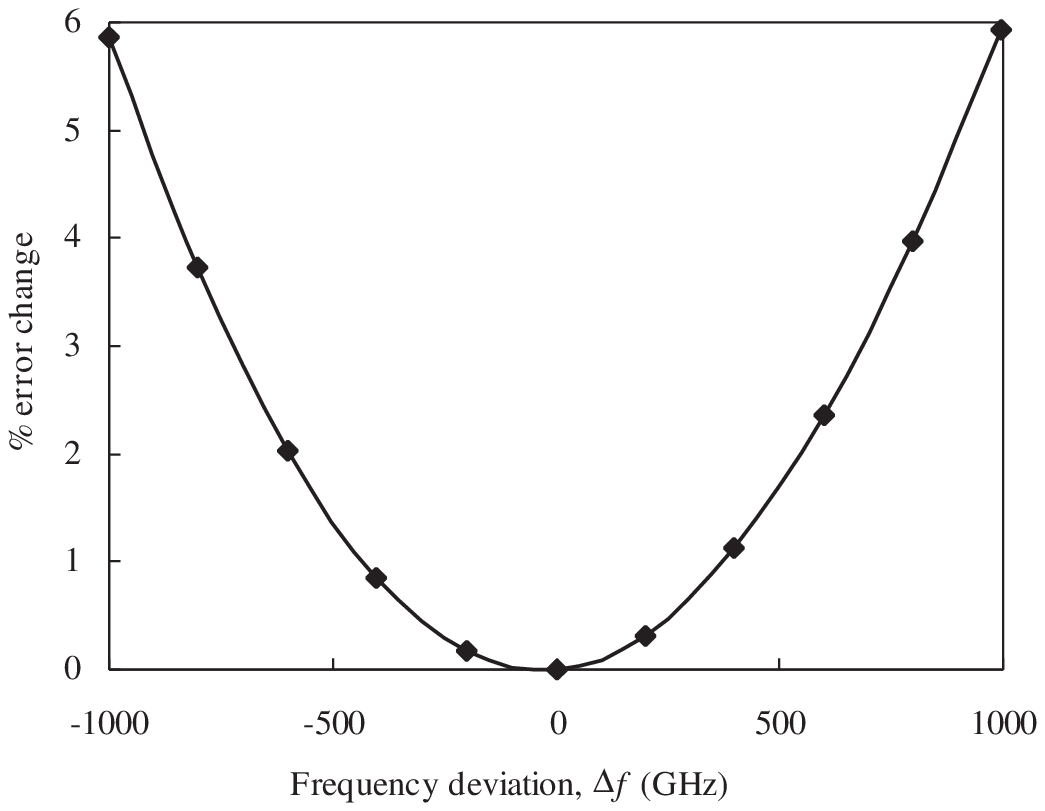}}
\vspace{7 cm}
 \\ Fig. 7
 \\ Gheorma, Haas, Levi

 \clearpage
\resizebox{15cm}{!}{\includegraphics{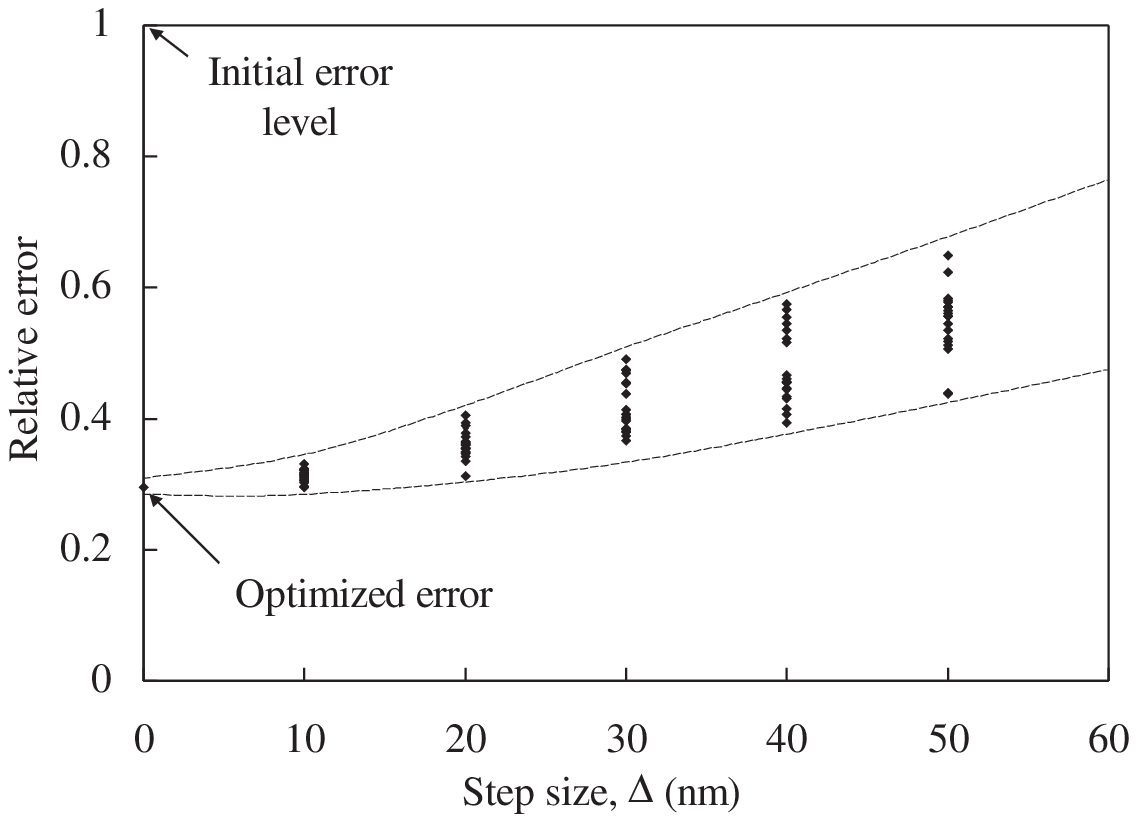}}
\vspace{7 cm}
 \\ Fig. 8
 \\ Gheorma, Haas, Levi


\begin{thebibliography}{99}
\bibitem{foot1} The problem considered here differs somewhat from
conventional inverse problems which usually involve calculations
on experimentally measured data.
\end{thebibliography}
\end{document}